# Mean Square Behavior of Noise-Robust Normalized Subband Adaptive Filter Algorithm

Yi Yu[1], Haiquan Zhao[2], Badong Chen[3], Wenyuan Wang[2], Lu Lu[4]

1 School of Information Engineering, Southwest University of Science and Technology, Mianyang, China
2 School of Electrical Engineering, Southwest Jiaotong University, Chengdu, China
3 School of Electronic and Information Engineering, Xi'an Jiaotong University, Xi'an, China
4 School of Electronics and Information Engineering, Sichuan University, Chengdu, China

Abstract: This paper studies the statistical models of the noise-robust normalized subband adaptive filter (NR-NSAF) algorithm in the mean and mean square deviation senses involving transient and steady-state behavior by resorting to the vectorization and Kronecker product of matrices. Thus, the proposed analysis does not require the Gaussian assumption to the input signal. Moreover, it removes the paraunitary assumption aiming to the analysis filter banks as in the existing analyses of subband adaptive algorithms. Simulation results in various conditions demonstrate the effectiveness of our theoretical analysis. For a special form of the algorithm, the proposed steady-state expression is also better accurate than the previous analysis.

## 1. Introduction

Adaptive filter algorithms have a pivotal position in some applications such as system identification, channel equalization, active noise control, and echo cancellation [1], [2]. One of the popular algorithms is the normalized least mean square (NLMS), and it is simple and easy in implementation. Nevertheless, the problem is very slow convergence rate for the correlated input signal. To overcome this problem, subband adaptive filter (SAF) with multiband structure has attracted much attention due to its decorrelation property. The multiband structure eliminates aliasing and band edge effects relative to the conventional structure [3]. According to this, the normalized SAF (NSAF) algorithm was proposed by Lee and Gan in [3]. Compared with the NLMS, the NSAF has faster convergence rate when the input signal has high correlation in the time-domain, while retaining comparable computational complexity. In a recent decade, many works have been reported to further obtain an improvement on the performance of the NSAF [4]-[12]. Typically, inspired by the NLMS with reusing weight vectors at each update [13], a noise-robust NSAF (NR-NSAF) algorithm was proposed which improves the steady-state performance in highly noisy environments [8], and almost at the same time, Ni proposed an improved NSAF (INSAF) algorithm [9]. And, the NR-NSAF algorithm is a more general form of the INSAF algorithm.

The performance analysis is a crucial point in the study of adaptive filter algorithms [2], [14]-[18]. Much literature has addressed the performance analysis of the NSAF algorithm [19]-[23]. Specifically, the steady-state mean-square error (MSE) of the NSAF using a fixed step size and a fixed regularization parameter were studied in [19] and [20], respectively. In some applications, e.g., system identification and echo cancellation, adaptive filter estimates the impulse response of the underlying system, so studying the mean square deviation (MSD) performance of adaptive algorithms seems to be more appropriate than the MSE. In general, the MSE can also be obtained from the MSD through the autocovariance matrix of input vector. In [24], Jeong *et al.* analyzed the steady-state MSD of the INSAF algorithm and this analysis framework has also been extended to its under-modeling scenario [25] and affine projection variant [10]. The theoretical results coincides with the simulations, but the accuracy depends on large number of subbands and long adaptive filter. However, the transient behavior of the INSAF algorithm has not been studied.

In this paper, we analyze the MSD performance of the NR-NSAF algorithm. Our analysis is based on the method of the vectorization and Kronecker product of matrices developed originally by Sayed [2]. This method is very popular recently, since it does not enforce the input signal to following a specified model (e.g., Gaussian distribution). Our contributions are summarized as below: 1) analyzing the transient and steady-state MSD of the NR-NSAF algorithm; 2) providing the mean condition on the step-size to ensure the algorithm stability. Moreover, unlike the existing analyses of subband adaptive algorithms, the proposed analysis does not assume the analysis filter banks to being paraunitary. Extensive simulations verify the proposed theoretical results.

Notations: $(\cdot)^T$, $\|\cdot\|_2$, $\lambda_{\max}(\cdot)$, $E\{\cdot\}$, $\mathrm{Tr}(\cdot)$, $\rho(\cdot)$, and $\otimes$ denote the transpose operator, the Euclidean norm of a vector, the largest eigenvalue of a matrix, the expectation of random variables, the trace of a matrix, the spectral radius of its matrix argument, and the Kronecker product, respectively. The notation $\mathrm{diag}\{\cdot\}$ yields the diagonal matrix according to its vector argument. The vectorization operator $\mathrm{vec}(\cdot)$ transforms an $M\times M$ matrix into an $M^2\times 1$ column vector by stacking successively the columns of the matrix, and $\mathrm{vec}^{-1}(\cdot)$ is the inverse of $\mathrm{vec}(\cdot)$. The symbols $\boldsymbol{I}$ and $\boldsymbol{0}$ denote the identity and zero matrices with appropriate sizes, respectively. Also, all the vectors in this paper are column vectors.

## 2. The NR-NSAF Algorithm

Supposing that the desired signal $d(n)$ is given by the linear model with respect to the input signal $u(n)$:

$$d(n) = \boldsymbol{u}^T(n)\boldsymbol{w}_o + \eta(n), \tag{1}$$

where $\boldsymbol{w}_o$ is an unknown $M$-length vector that needs to be identified, $\boldsymbol{u}(n) = [u(n), u(n-1), ..., u(n-M+1)]^T$ is the input vector, and $\eta(n)$ is the system noise. Fig. 1 shows the multiband-structured SAF with $N$ subbands. The input signal $u(n)$ and the desired signal $d(n)$ are partitioned into multiple subband signals $u_i(n)$ and $d_i(n)$ by the analysis filters $H_i(z)$, $i = 0,1,...,N-1$, respectively. The subband outputs $y_i(n)$ are obtained by filtering signals $u_i(n)$ through a fullband adaptive filter whose weight vector is denoted as $\boldsymbol{w}(k) = [w_1(k),\ w_2(k),\ ...,\ w_M(k)]^T$. Then, by $N$-fold decimating signals $y_i(n)$ and $d_i(n)$, we can obtain $y_{i,D}(k)$ and $d_{i,D}(k)$ at lower sampling rate, respectively. We use $n$ and $k$ to represent the original sequences and the decimated sequences, respectively. For the $i$-th subband, the decimated error signal is expressed as

$$e_{i,\mathrm{D}} = d_{i,D}(k) - \boldsymbol{u}_i^T(k)\boldsymbol{w}(k), \tag{2}$$

where $\boldsymbol{u}_i(k) = [u_i(kN),\ u_i(kN-1),\ ...,\ u_i(kN-M+1)]^T$ and $d_{i,D}(k) = d_i(kN)$.

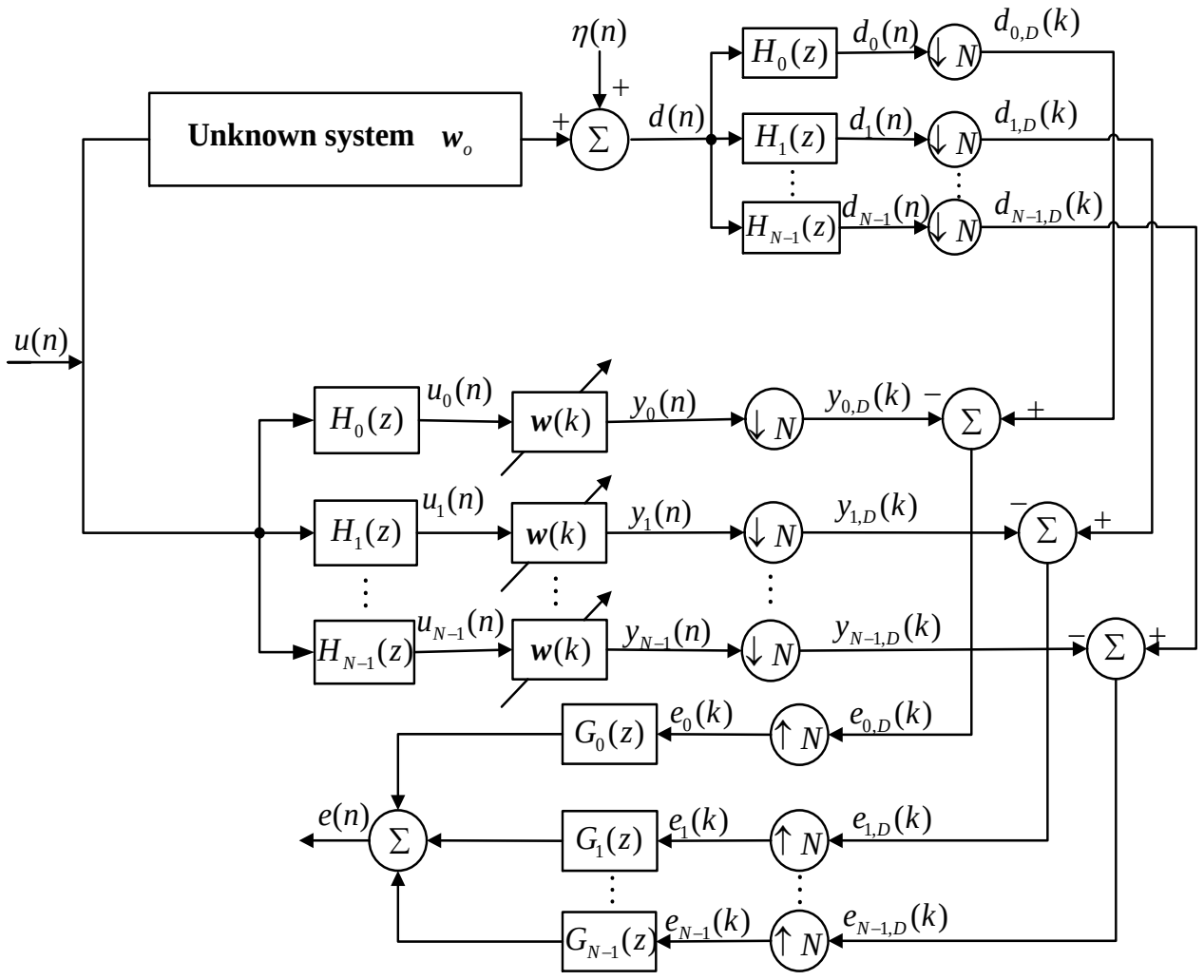

Fig. 1 Multiband structured SAF.

In [8], the NR-NSAF algorithm for updating the weight vector is described as

$$\boldsymbol{w}(k+1) = \sum_{p=0}^{P-1}\beta_p \boldsymbol{w}(k-p) + \mu\sum_{i=0}^{N-1}\frac{\zeta_{i,\mathrm{D}}(k)\boldsymbol{u}_i(k)}{\|\boldsymbol{u}_i(k)\|_2^2 + \varepsilon}, \tag{3}$$

$$\zeta_{i,\mathrm{D}} = d_{i,D}(k) - \boldsymbol{u}_i^T(k)\sum_{p=0}^{P-1}\beta_p \boldsymbol{w}(k-p), \tag{4}$$

where $\beta_p = \left(\sum_{p=0}^{P-1}\alpha^p\right)^{-1}\alpha^p$, $\alpha$ ($0 < \alpha \le 1$) is a weighting factor, $\mu > 0$ is the step-size, $\varepsilon > 0$ is a small regularization constant to avoid the division by zero, and $P$ denotes the number of reusing recent weight vectors at each iteration. Note that, the NR-NSAF algorithm reduces to the INSAF and NSAF algorithms when $\alpha = 1$ and $P = 1$, respectively.

## 3. Performance analysis

Let us introduce two matrices: $\boldsymbol{U}(k) = \left[\boldsymbol{u}(kN), \boldsymbol{u}(kN-1), ..., \boldsymbol{u}(kN-L+1)\right]$, $\boldsymbol{H} = \left[\boldsymbol{h}_0, \boldsymbol{h}_1, ..., \boldsymbol{h}_{N-1}\right]$, and two vectors: $\boldsymbol{d}(k) = \left[d(kN), d(kN-1), ..., d(kN-L+1)\right]^T$, $\boldsymbol{u}(kN) = [u(kN),\ u(kN-1),\ ...,\ u(kN-M+1)]^T$, where $\boldsymbol{h}_i$ is the impulse response of the $i$-th analysis filter $H_i(z)$ with length of $L$. Then, we can find the following relations:

$$\boldsymbol{U}_D(k) \triangleq \left[\boldsymbol{u}_0(k), \boldsymbol{u}_1(k), ..., \boldsymbol{u}_{N-1}(k)\right] = \boldsymbol{U}(k)\boldsymbol{H}, \tag{5}$$

$$\boldsymbol{d}_D(k) \triangleq \left[d_{0,D}(k), d_{1,D}(k), ..., d_{N-1,D}(k)\right]^T = \boldsymbol{H}^T\boldsymbol{d}(k), \tag{6}$$

$$\boldsymbol{d}(k) = \boldsymbol{U}^T(k)\boldsymbol{w}_o + \boldsymbol{\eta}(k), \tag{7}$$

where $\boldsymbol{\eta}(k) = \left[\eta(kN), \eta(kN-1), ..., \eta(kN-L+1)\right]^T$. We use (4)-(6) to rearrange (3) as

$$\begin{aligned}\boldsymbol{w}(k+1) = &\sum_{p=0}^{P-1}\beta_p \boldsymbol{w}(k-p) + \mu\boldsymbol{U}(k)\boldsymbol{H}\boldsymbol{\Lambda}^{-1}(k)\times \\ &\left[\boldsymbol{H}^T\boldsymbol{d}(k) - \boldsymbol{H}^T\boldsymbol{U}^T(k)\sum_{p=0}^{P-1}\beta_p \boldsymbol{w}(k-p)\right],\end{aligned} \tag{8}$$

where $\boldsymbol{\Lambda}(k) = \varepsilon\boldsymbol{I}_N + \mathrm{diag}\left\{\boldsymbol{U}_D^T(k)\boldsymbol{U}_D(k)\right\}$.

Subtracting $\boldsymbol{w}_o$ from both sides of (8) yields

$$\tilde{\boldsymbol{w}}(k+1) = \left[\boldsymbol{I}_M - \mu\boldsymbol{A}(k)\right]\sum_{p=0}^{P-1}\beta_p\tilde{\boldsymbol{w}}(k-p) - \mu\boldsymbol{b}(k), \tag{9}$$

where $\tilde{\boldsymbol{w}}(k) \triangleq \boldsymbol{w}_o - \boldsymbol{w}(k)$ represents the weight error vector as, $\boldsymbol{A}(k) = \boldsymbol{U}_D(k)\boldsymbol{\Lambda}^{-1}(k)\boldsymbol{U}_D^T(k)$ and $\boldsymbol{b}(k) = \boldsymbol{U}_D(k)\boldsymbol{\Lambda}^{-1}(k)\boldsymbol{H}^T\boldsymbol{\eta}(k)$. For deriving (9), we also use the relation $\sum_{p=0}^{P-1}\beta_p = 1$.

Before proceeding, we define some block matrices and vectors:
$\boldsymbol{\mathcal{A}}(k) = \begin{bmatrix}\boldsymbol{A}(k) & \boldsymbol{0}_{M\times M(P-1)} \\ \boldsymbol{0}_{M(P-1)\times M} & \boldsymbol{0}_{M(P-1)\times M(P-1)}\end{bmatrix}$,
$\boldsymbol{\mathcal{C}} = \begin{bmatrix} & \boldsymbol{\beta} \\ \boldsymbol{I}_{M(P-1)} & \boldsymbol{0}_{M(P-1)\times M}\end{bmatrix}$, $\tilde{\boldsymbol{W}}(k) = \begin{bmatrix}\tilde{\boldsymbol{w}}(k) \\ \vdots \\ \tilde{\boldsymbol{w}}(k-P+1)\end{bmatrix}$,
$\boldsymbol{B}(k) = \begin{bmatrix}\boldsymbol{b}(k) \\ \boldsymbol{0}_{M(P-1)\times 1}\end{bmatrix}$, where $\boldsymbol{\beta} = \left[\beta_0, ..., \beta_{P-1}\right]\otimes\boldsymbol{I}_M$. With these definitions, (9) can be rewritten as

$$\tilde{\boldsymbol{W}}(k+1) = \left[\boldsymbol{I}_{MP} - \mu\boldsymbol{\mathcal{A}}(k)\right]\boldsymbol{\mathcal{C}}\tilde{\boldsymbol{W}}(k) - \mu\boldsymbol{B}(k). \tag{10}$$

Equation (10) will be the starting point to analyze the performance of the NR-NSAF algorithm.

For tractable analysis, the following assumptions are necessary.

A1): The system noise $\eta(n)$ is a white process with zero-mean and variance $\sigma_\eta^2$, which is independent of $u(n)$.

A2): $\boldsymbol{u}(n)$ is zero-mean stationary random vector with positive definite covariance matrix.

A3): $\boldsymbol{u}_i(k)$ for $i=0,1,...,N-1$ and $\boldsymbol{w}(k-p)$ for $p=0,...,P-1$ are independent each other, which is the well-known *independence assumption* used for analyzing the performance of adaptive algorithms [2], [14]-[27].

From the assumption A2), $\boldsymbol{u}_i(k)$ for $i=0,1,...,N-1$ are also zero-mean stationary with positive definite covariance matrices. According to assumptions A1) and A3), we can further assume that $\tilde{\boldsymbol{W}}(k)$ is independent of $\boldsymbol{\mathcal{A}}(k)$ and $\boldsymbol{B}(k)$.

*3.1 Mean behavior*

Under assumptions A1) and A3), the expectation of both sides of (10) is obtained:

$$E\left\{\tilde{\boldsymbol{W}}(k+1)\right\}=\left(\boldsymbol{I}_{MP}-\mu E\left\{\boldsymbol{\mathcal{A}}(k)\right\}\right)\boldsymbol{\mathcal{C}}E\left\{\tilde{\boldsymbol{W}}(k)\right\}. \quad (11)$$

*Theorem 1*: The NR-NSAF algorithm is mean stable if, and only if the step size satisfy

$$0<\mu<\frac{2}{\lambda_{\max}\left(E\left\{\boldsymbol{A}(k)\right\}\right)}. \quad (12)$$

*Proof*: See Appendix.

At the steady-state, i.e., when $k\to\infty$, we obtain from (11):

$$E\left\{\tilde{\boldsymbol{W}}(\infty)\right\}=\boldsymbol{0}_{MP\times 1}. \quad (13)$$

The above relation means that the NR-NSAF algorithm can yield an unbiased estimate for the unknown vector $\boldsymbol{w}_o$.

*3.2 Mean square behavior*

Post-multiplying (10) with its transpose and defining $\boldsymbol{\Phi}(k)\triangleq\tilde{\boldsymbol{W}}(k)\tilde{\boldsymbol{W}}^T(k)$, the following matrix recursion is developed:

$$\begin{aligned}\boldsymbol{\Phi}(k+1)=&\boldsymbol{\mathcal{C}}\boldsymbol{\Phi}(k)\boldsymbol{\mathcal{C}}^T-\mu\boldsymbol{\mathcal{A}}(k)\boldsymbol{\mathcal{C}}\boldsymbol{\Phi}(k)\boldsymbol{\mathcal{C}}^T\\ &-\mu\boldsymbol{\mathcal{C}}\boldsymbol{\Phi}(k)\boldsymbol{\mathcal{C}}^T\boldsymbol{\mathcal{A}}^T(k)\\ &+\mu^2\boldsymbol{\mathcal{A}}(k)\boldsymbol{\mathcal{C}}\boldsymbol{\Phi}(k)\boldsymbol{\mathcal{C}}^T\boldsymbol{\mathcal{A}}^T(k)\\ &+\mu^2\boldsymbol{B}(k)\boldsymbol{B}^T(k)\\ &-\mu\left[\boldsymbol{I}_{MP}-\mu\boldsymbol{\mathcal{A}}(k)\right]\boldsymbol{\mathcal{C}}\tilde{\boldsymbol{W}}(k)\boldsymbol{B}^T(k)\\ &-\mu\boldsymbol{B}(k)\tilde{\boldsymbol{W}}^T(k)\boldsymbol{\mathcal{C}}^T\left[\boldsymbol{I}_{MP}-\mu\boldsymbol{\mathcal{A}}(k)\right]^T.\end{aligned} \quad (14)$$

With assumptions A1-A3), the expectations of the last two terms in (14) are zero. Consequently, enforcing the expectation operator on both sides of (14), we have

$$\begin{aligned}E\left\{\boldsymbol{\Phi}(k+1)\right\}=&\boldsymbol{\mathcal{C}}E\left\{\boldsymbol{\Phi}(k)\right\}\boldsymbol{\mathcal{C}}^T\\ &-\mu E\left\{\boldsymbol{\mathcal{A}}(k)\right\}\boldsymbol{\mathcal{C}}E\left\{\boldsymbol{\Phi}(k)\right\}\boldsymbol{\mathcal{C}}^T\\ &-\mu\boldsymbol{\mathcal{C}}E\left\{\boldsymbol{\Phi}(k)\right\}\boldsymbol{\mathcal{C}}^TE\left\{\boldsymbol{\mathcal{A}}^T(k)\right\}\\ &+\mu^2E\left\{\boldsymbol{\mathcal{A}}(k)\boldsymbol{\mathcal{C}}E\left\{\boldsymbol{\Phi}(k)\right\}\boldsymbol{\mathcal{C}}^T\boldsymbol{\mathcal{A}}^T(k)\right\}\\ &+\mu^2E\left\{\boldsymbol{B}(k)\boldsymbol{B}^T(k)\right\}.\end{aligned} \quad (15)$$

The last expectation term of (15) can be further expressed as

$$E\left\{\boldsymbol{B}(k)\boldsymbol{B}^T(k)\right\}=\begin{bmatrix}E\left\{\boldsymbol{b}(k)\boldsymbol{b}^T(k)\right\} & \boldsymbol{0}_{M\times M(P-1)}\\ \boldsymbol{0}_{M(P-1)\times M} & \boldsymbol{0}_{M(P-1)\times M(P-1)}\end{bmatrix}, \quad (16)$$

$$\begin{aligned}E\left\{\boldsymbol{b}(k)\boldsymbol{b}^T(k)\right\}&=\sigma_\eta^2E\left\{\boldsymbol{U}_D(k)\boldsymbol{\Lambda}^{-1}(k)\boldsymbol{H}^T\boldsymbol{H}\boldsymbol{\Lambda}^{-1}(k)\boldsymbol{U}_D^T(k)\right\}\\ &=\sigma_\eta^2\sum_{i=0}^{N-1}\|\boldsymbol{h}_i\|_2^2E\left\{\frac{\boldsymbol{u}_i(k)\boldsymbol{u}_i^T(k)}{\left(\|\boldsymbol{u}_i(k)\|_2^2+\varepsilon\right)^2}\right\}\end{aligned}. \quad (17)$$

To continue the analysis, we need two properties of the Kronecker product, namely,

$$\text{vec}\left(\boldsymbol{XZY}\right)=\left(\boldsymbol{Y}^T\otimes\boldsymbol{X}\right)\text{vec}\left(\boldsymbol{Z}\right) \quad (18)$$

and

$$\boldsymbol{XY}\otimes\boldsymbol{Z\Omega}=\left(\boldsymbol{X}\otimes\boldsymbol{Z}\right)\left(\boldsymbol{Y}\otimes\boldsymbol{\Omega}\right) \quad (19)$$

for any matrices $\left\{\boldsymbol{X},\boldsymbol{Y},\boldsymbol{Z},\boldsymbol{\Omega}\right\}$ of compatible dimensions [28]. Subsequently, taking the vectorization for all the matrices in the recursion (15), after merging similar terms, it is established that

$$\begin{aligned}\text{vec}\left(E\left\{\boldsymbol{\Phi}(k+1)\right\}\right)=&\boldsymbol{F}\text{vec}\left(E\left\{\boldsymbol{\Phi}(k)\right\}\right)\\ &+\mu^2\text{vec}\left(E\left\{\boldsymbol{B}(k)\boldsymbol{B}^T(k)\right\}\right),\end{aligned} \quad (20)$$

where the $M^2P^2\times M^2P^2$ matrix $\boldsymbol{F}$ is given by

$$\begin{aligned}\boldsymbol{F}=&\left[\boldsymbol{I}_{M^2P^2}-\mu\left(\boldsymbol{I}_{MP}\otimes E\left\{\boldsymbol{\mathcal{A}}(k)\right\}\right)-\mu\left(E\left\{\boldsymbol{\mathcal{A}}(k)\right\}\otimes\boldsymbol{I}_{MP}\right)\right.\\ &\left.+\mu^2E\left\{\boldsymbol{\mathcal{A}}(k)\otimes\boldsymbol{\mathcal{A}}(k)\right\}\right]\left(\boldsymbol{\mathcal{C}}\otimes\boldsymbol{\mathcal{C}}\right).\end{aligned} \quad (21)$$

The MSD is defined as

$$\begin{aligned}\text{MSD}(k)\triangleq&E\left\{\|\tilde{\boldsymbol{w}}(k)\|_2^2\right\}=\text{Tr}\left\{E\left\{\tilde{\boldsymbol{w}}(k)\tilde{\boldsymbol{w}}^T(k)\right\}\right\}\\ &=\text{Tr}\left\{\left[E\left\{\boldsymbol{\Phi}(k)\right\}\right]_1\right\}.\end{aligned} \quad (22)$$

where $\left[E\left\{\boldsymbol{\Phi}(k)\right\}\right]_1$ is the first $M\times M$ diagonal block of $E\left\{\boldsymbol{\Phi}(k)\right\}$. So, based on the inverse operator $\text{vec}^{-1}(\cdot)$, the recursion (20) models the MSD evolution behavior of the NR-NSAF algorithm with respect to the iteration *k*.

It is seen from (21) that the NR-NSAF algorithm is convergent in the mean square sense if, and only if, the matrix $\boldsymbol{F}$ is stable that all the eigenvalues of $\boldsymbol{F}$ are in the range $(-1,\ 1)$ [14]. However, further obtaining the step size range from it is difficult due to the existence of the matrix $\left(\boldsymbol{\mathcal{C}}\otimes\boldsymbol{\mathcal{C}}\right)$. Fortunately, we have deduced the mean square convergence condition $0<\mu<2$ in an alternative method, see Appendix in [12].

Then, if the algorithm converges to the steady-state, the equality $E\left\{\boldsymbol{\Phi}(k+1)\right\}=E\left\{\boldsymbol{\Phi}(k)\right\}$ as $k\to\infty$ will be hold. Accordingly, we can arrive at the steady-state solution of (20):

$$E\left\{\boldsymbol{\Phi}(\infty)\right\}=\mu^2\text{vec}^{-1}\left(\left(\boldsymbol{I}_{M^2P^2}-\boldsymbol{F}\right)^{-1}\text{vec}\left(E\left\{\boldsymbol{B}(k)\boldsymbol{B}^T(k)\right\}\right)\right). \quad (23)$$

Using the relation $\text{Tr}\left(\boldsymbol{XY}\right)=\left(\text{vec}(\boldsymbol{X}^T)\right)^T\text{vec}(\boldsymbol{Y})$, the steady-state MSD of the NR-NSAF algorithm can be derived from (23), i.e.,

$$\begin{aligned}\text{MSD}(\infty)=&\text{Tr}\left\{\left[E\left\{\boldsymbol{\Phi}(\infty)\right\}\right]_1\right\}/P\\ &=\mu^2\text{vec}^T(\boldsymbol{I}_{MP})\left(\boldsymbol{I}_{M^2P^2}-\boldsymbol{F}\right)^{-1}\times\\ &\text{vec}\left(E\left\{\boldsymbol{B}(k)\boldsymbol{B}^T(k)\right\}\right)/P.\end{aligned} \quad (24)$$

Remark 1. Reference [24] presents an MSD(∞) expression

for the simple form of the NR-NSAF algorithm when $\alpha$=1 and $\varepsilon$=0 (that is the INSAF algorithm). Nonetheless, it benefits from two extra assumptions: 1) when the number of subbands is sufficiently large, the decimated subband input signals are close to the white signals, i.e., $E\{\boldsymbol{u}_i(k)\boldsymbol{u}_i^T(k)\} \approx \sigma_{u_i}^2 \boldsymbol{I}_M$ and $E\{\boldsymbol{u}_i^T(k)\boldsymbol{u}_i(k)\} \approx M\sigma_{u_i}^2$; 2) The length $M$ of adaptive filter is large so that the fluctuation of the energy of the decimated subband input signals from one iteration to the next iteration is small. In addition, the MSD($\infty$) in [24] requires knowing the variances of the subband noises, $\sigma_{\eta_i}^2$, which is usually given by $\sigma_{\eta_i}^2 = \sigma_{\eta_i}^2 / N$ under the paraunitary assumption for the analysis filters [21]. However, we propose to use $\sigma_{\eta_i}^2 = \|\boldsymbol{h}_i\|_2^2 \sigma_\eta^2$.

## 4. Simulation results

Simulations are conducted in the system identification. Both the adaptive filter and the unknown vector have the same length $M$=16. The weight vector of adaptive filter is initialized as a null vector. The correlated input signal $u(n)$ is generated by filtering either a zero mean white Gaussian signal with unit variance or a uniform distribution signal with the interval $[-1, 1]$, through a first-order autoregressive system of a pole at 0.9 [22], called as the Gaussian input and the uniform input in simulations, respectively. The system noise $\eta(n)$ is a zero mean white Gaussian process, giving to a certain signal-to-noise rate (SNR). The analysis filters are designed based on the cosine modulated filter banks, where the length of the prototype filter is 64 when the number of subbands is $N$=8, unless otherwise specified. For evaluating the proposed theoretical expressions, the expectations on the subband inputs shown in (17) and (21) are estimated by ensemble averaging. The regularization constant $\varepsilon$ is set to 0.001 except Fig. 8. All the simulations results are averaging over 200 independent trials.

### *4.1 Transient performance*

To begin with, the mean evolution behavior of the algorithm is checked in Fig. 2 for identifying the unknown vector $\boldsymbol{w}_o$=[0.51, −0.04, 0.02, 0.09, 0.22, 0.20, 0.13, −0.48, −0.39, 0.32, −0.11, −0.30, 0.25, −0.24, 0.6, −0.01]$^T$. It is clear to see that the theoretical weights calculated by (11) match well with the simulated weights.

In the following examples, we investigate the MSD performance of the algorithm by $10\log_{10}\text{MSD}(k)$ (dB). The unknown vector is randomly generated by using the function $rand(M,1)-0.5$ in MATLAB and normalized by $\boldsymbol{w}_o^T\boldsymbol{w}_o$=1. Fig. 3 shows the effect of the parameter $\alpha$ on the NR-NSAF performance. As can be seen, the theoretical MSD curves computed by (20) have good agreement with the simulated curves. In addition, for values of $\alpha$ closer to 1, the steady-state MSD is lower while retaining comparable convergence rate. Therefore, $\alpha = 1$ is preferred for the NR-NSAF algorithm.

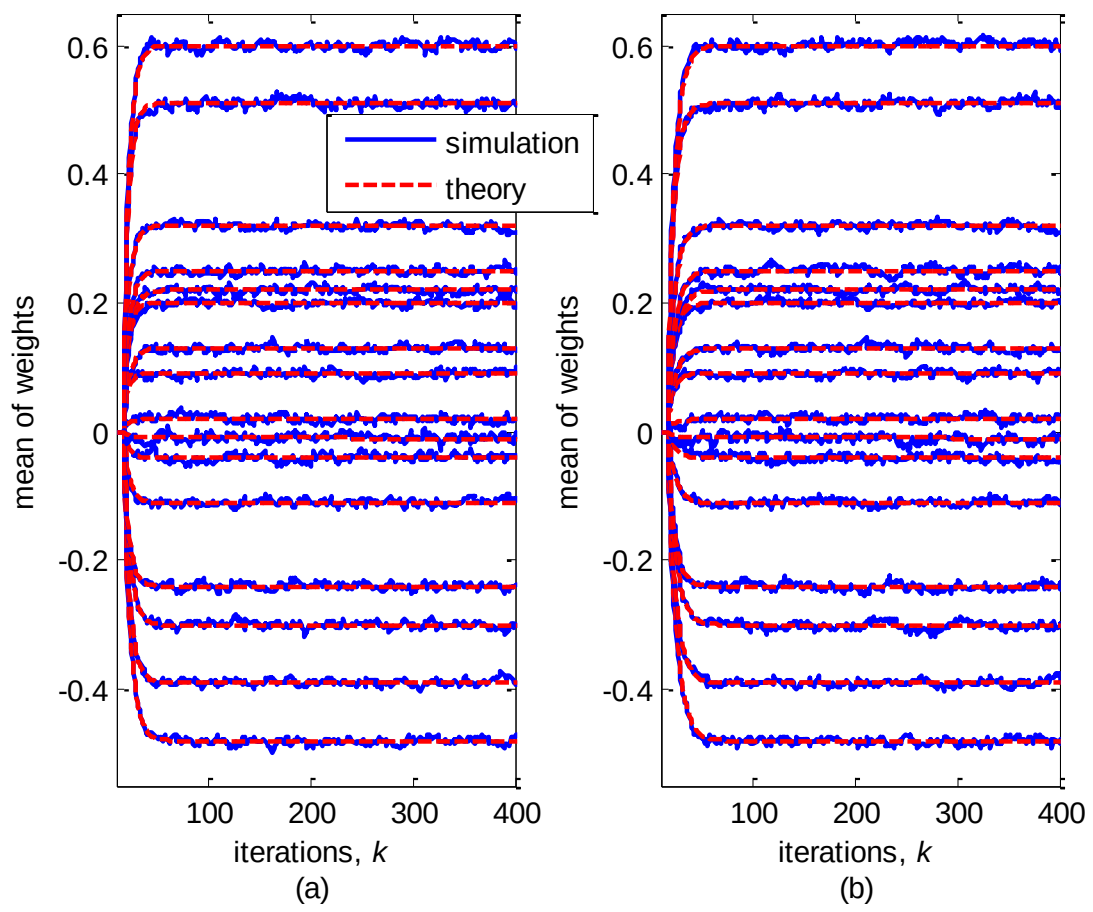

Fig. 2 Mean behavior of the algorithm. (a) $\alpha = 0.5$, (b) $\alpha = 1$. [Gaussian input, SNR=10 dB, $\mu$=0.5, and $P$=3].

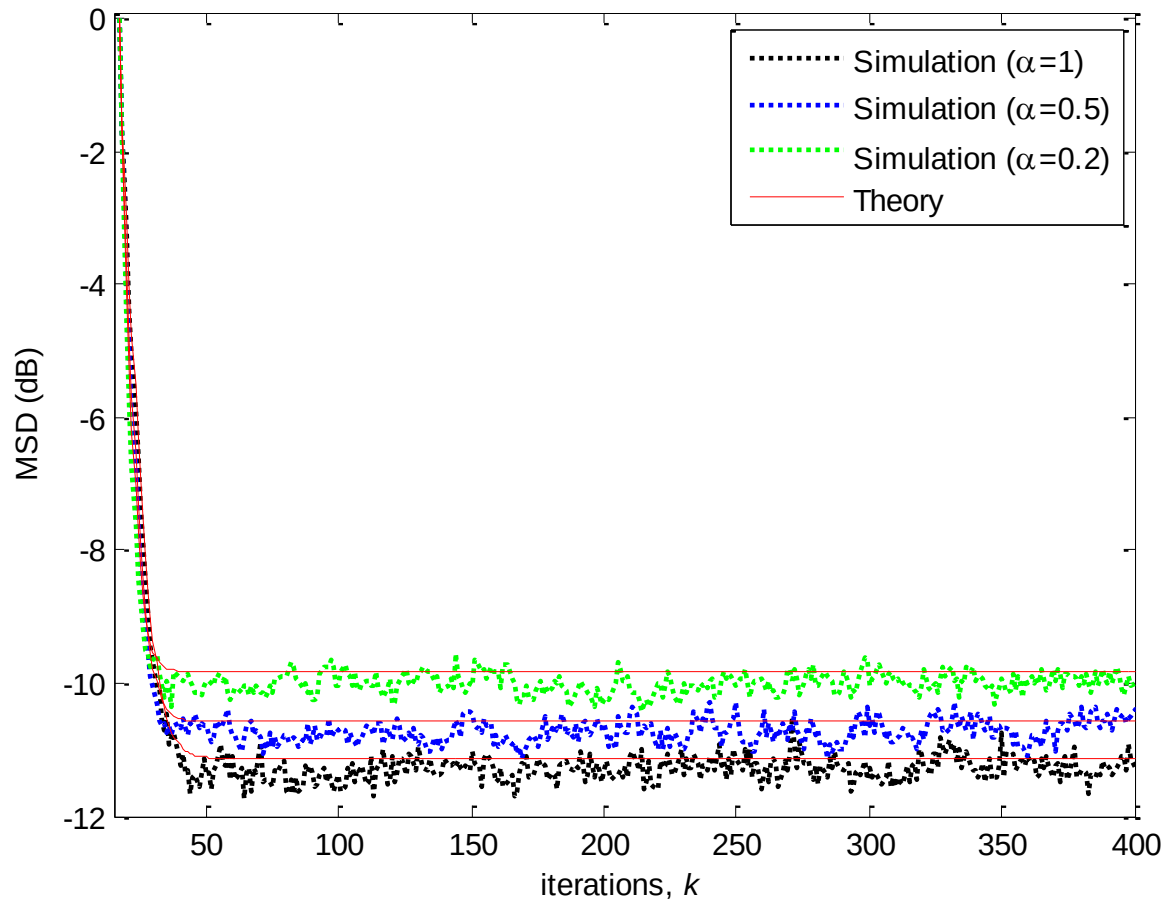

Fig. 3 MSD curves of the algorithm versus $\alpha$ values. [Gaussian input, SNR = 10 dB, $\mu = 0.5$ and $P = 3$].

Fig. 4 depicts the MSD results of the NR-NSAF algorithm with $P$=1, 2, 3 and 4 values, where $P$=1 also denotes the NSAF algorithm. It is seen that theoretical calculation gives good fit with the simulation. Fig. 4(a) also shows that, in a low SNR scenario, the NR-NSAF algorithm has smaller steady-state MSD for larger $P$, without slowing convergence. In a high SNR case see Fig. 4(b), however, the convergence of the algorithm is slowed as $P$ increases. It follows that the NR-NSAF algorithm will work better than the NSAF algorithm when the environment is highly noisy.

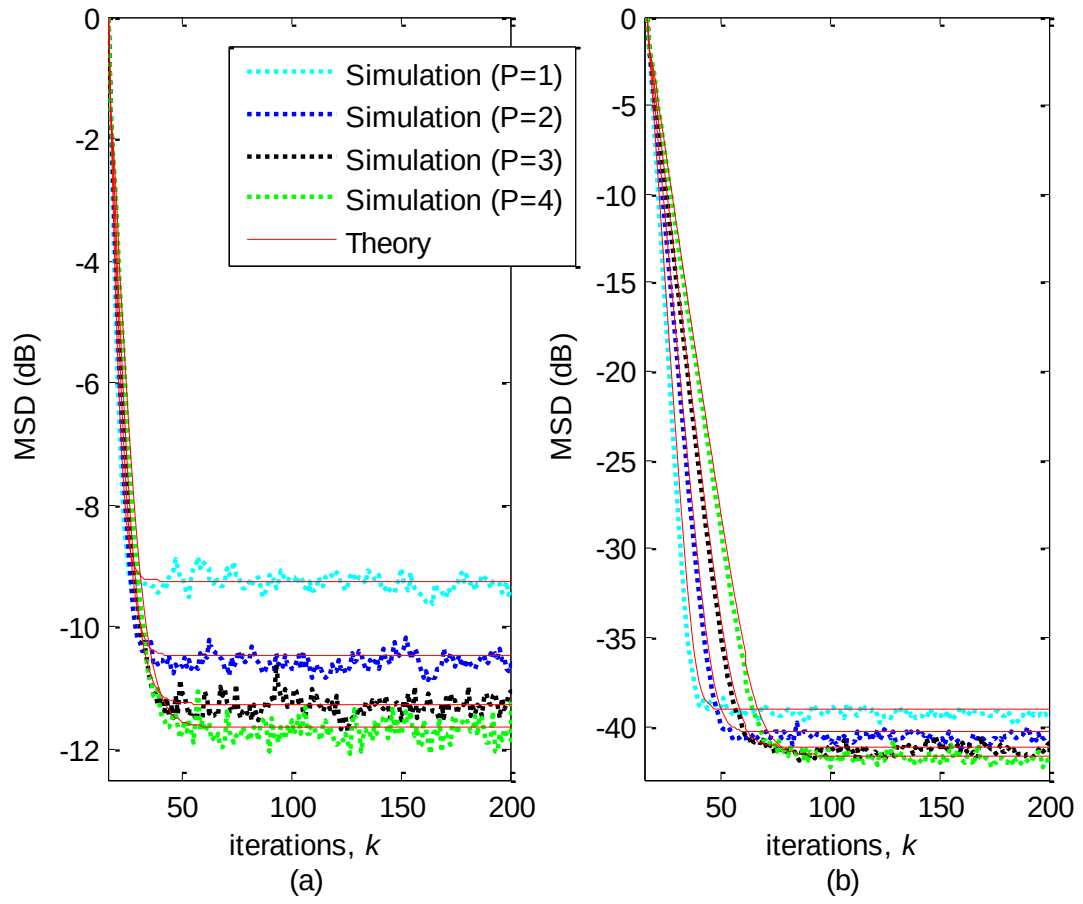

Fig. 4 MSD curves of the algorithm versus $P$ values. (a) SNR=10 dB, (b)

SNR=40 dB. [ $\mu=0.5$ and $\alpha=1$ ].

In Fig. 5, the MSD performance curves of the NR-NSAF algorithm with different step sizes ($\mu$=0.5, 0.1 and 0.4) are shown for both Gaussian and uniform inputs. Fig. 6 depicts similar results when the number of subbands is $N$=4. As one can see, the theoretical results fairly agree with the simulated results. It is worth noting that, as the member of constant step size based adaptive algorithms, users need to consider a compromise between fast convergence and low steady-state MSD when choosing the step size for the NR-NSAF algorithm.

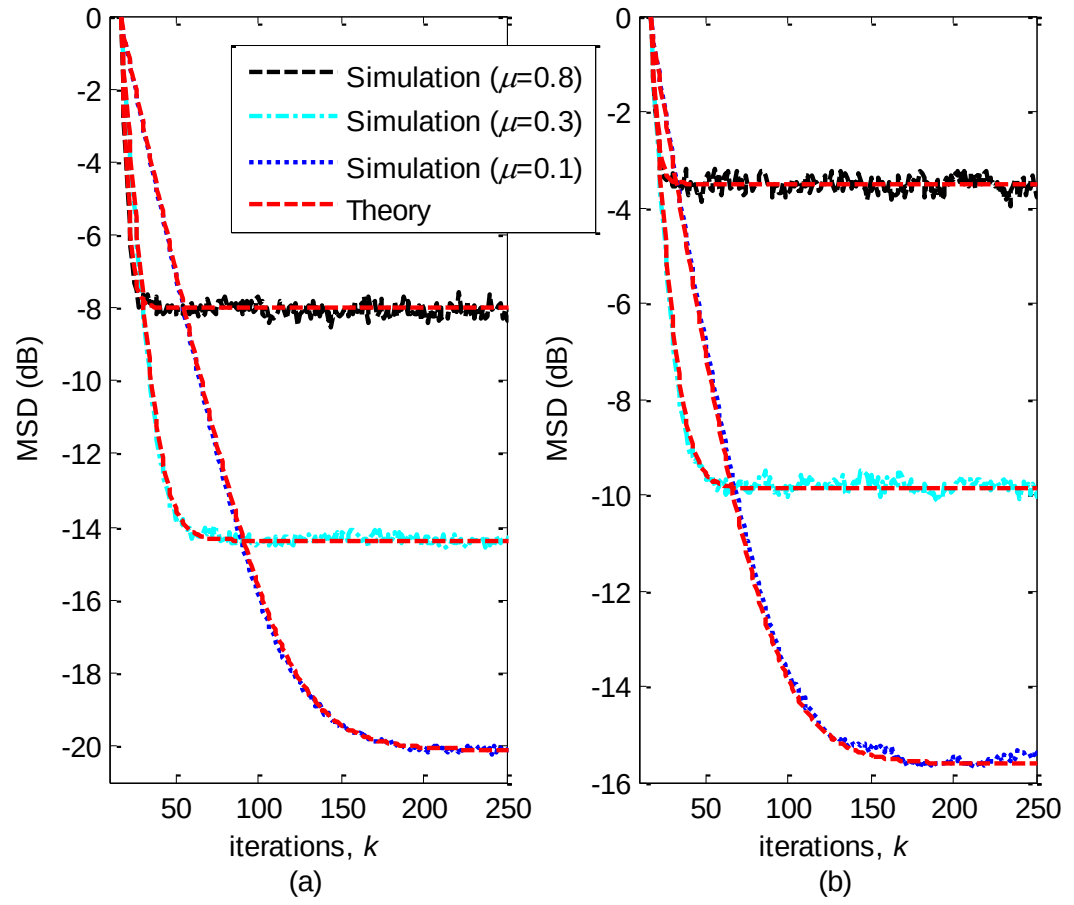

Fig. 5 MSD curves of the algorithm versus step sizes. (a) Gaussian input, (b) uniform input. [ $\text{SNR}=10$ dB , $N$=8, $P=3$ , and $\alpha=1$ ].

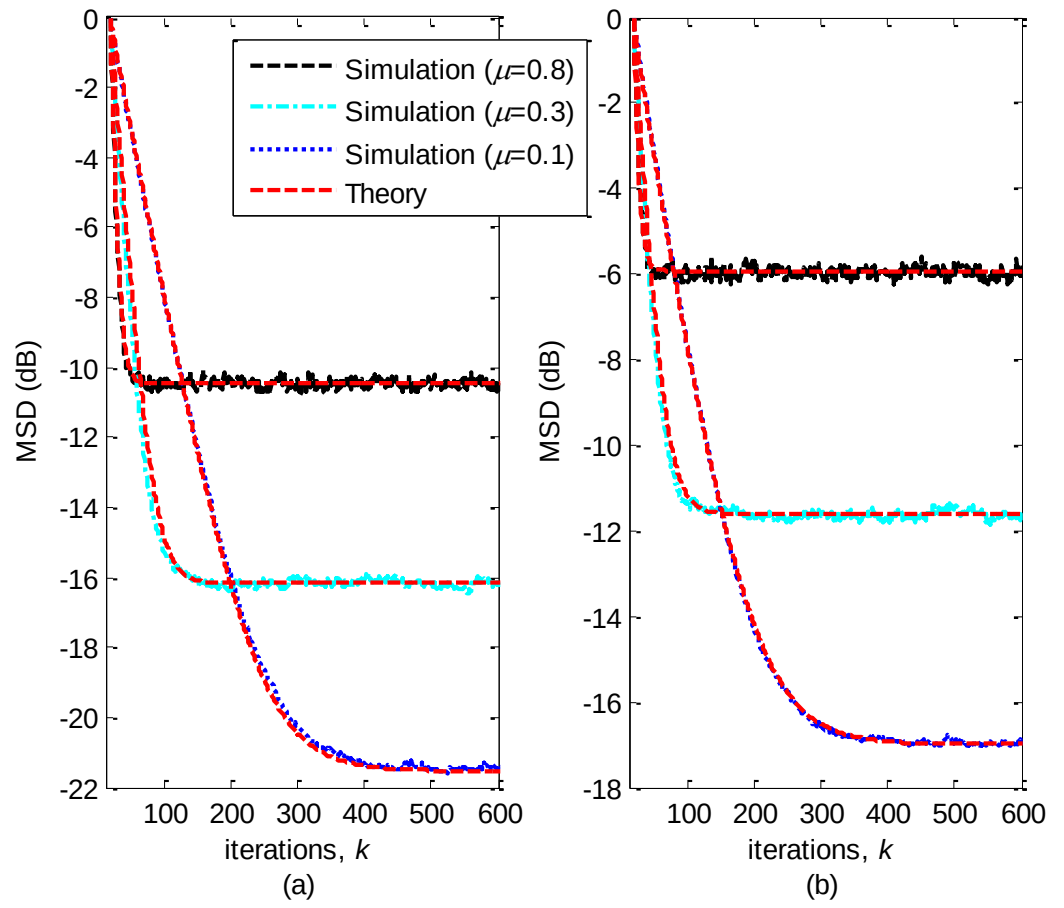

Fig. 6 MSD curves of the algorithm versus step sizes. (a) Gaussian input, (b) uniform input. [ $\text{SNR}=10$ dB , $N$=4, $P=3$ , and $\alpha=1$ ].

### *4.2 Steady-state performance*

Fig. 7 examines the effectiveness of (22) for predicting the steady-state MSD of the NR-NSAF algorithm as a function of the step size. The simulation values are the average over 200 MSDs at the steady-state stage. The step size $\mu$ is increased from 0.1 to 1. To fairly comparing with the theory presented in [24], we choose $\alpha=1$ and $\varepsilon=0$. As can be seen from Fig. 7, the proposed theoretical results have a good match with the simulated results for a smaller step size. However, a discrepancy of them can also be observed in Fig. 7 for larger step sizes. In comparison, the proposed (38) works better than the theory from [24] used for predicting the steady-state MSD of the NR-NSAF algorithm. This is because that the model in [24] requires large enough number of subbands and long adaptive filter.

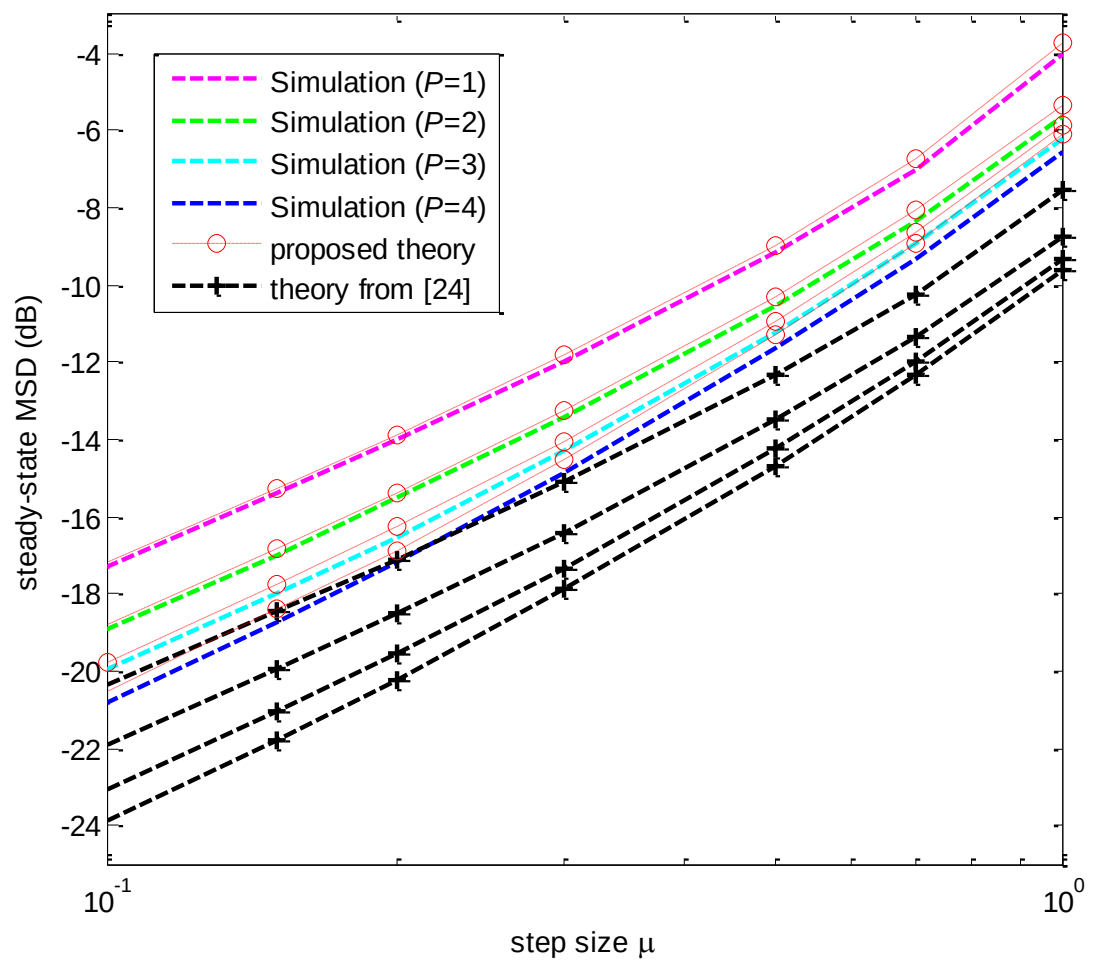

Fig. 7 Steady-state MSDs versus step sizes. [Gaussian input].

## 5. Conclusion

We have analyzed in detail the performance of the NR-NSAF algorithm in terms of the transient and steady-state MSD. The proposed analysis is based on the vectorization and the Kronecker product of matrices, thereby it drops the specified distribution assumption of the input signal. In addition, the paraunitary assumption is unnecessary for the analysis filter banks in our analysis. For the special INSAF algorithm, the proposed steady-state expression outperforms the previous theory in [24] for a low-order adaptive filter scenario. Simulation results have shown good agreement with the theoretical results.

## Appendix

In order to ensure the stability that the recursion (11) evolves with the iteration $k$, all the eigenvalues of the matrix $\boldsymbol{\Xi} \triangleq \left(\boldsymbol{I}_{MP} - \mu E\{\boldsymbol{\mathcal{A}}(k)\}\right)\boldsymbol{\mathcal{C}}$ must be inside the unit circle, i.e., $\rho(\boldsymbol{\Xi})<1$. We rearrange $\boldsymbol{\Xi}$ as

$$\boldsymbol{\Xi} \triangleq \begin{bmatrix} \boldsymbol{\Xi}_0, \ \boldsymbol{\Xi}_1 & \cdots & \boldsymbol{\Xi}_{P-1} \\ \boldsymbol{I}_{M(P-1)} & & \boldsymbol{0}_{M(P-1)\times M} \end{bmatrix}, \tag{31}$$

where $\boldsymbol{\Xi}_p = \beta_p\left(\boldsymbol{I}_M - \mu E\{\mathbf{A}(k)\}\right)$ for $p=0,1,...,P-1$ . To proceed, we take advantage of the block-maximum-norm of the block matrix, with the notation $\|\cdot\|_{b\infty}$ [29]:

$$\begin{aligned} \|\boldsymbol{\Psi}\|_{b\infty} &\triangleq \max_{\boldsymbol{x}\neq\boldsymbol{0}} \|\boldsymbol{\Psi x}\|_{b\infty} / \|\boldsymbol{x}\|_{b\infty} , \\ \|\boldsymbol{x}\|_{b\infty} &\triangleq \max_{0\le p\le P-1} \|\boldsymbol{x}_p\|_2 , \end{aligned} \tag{32}$$

where $\boldsymbol{\Psi}$ is an $MP\times MP$ matrix with block entries of size $M\times M$ each, and $\boldsymbol{x}=\left[\boldsymbol{x}_0^T,...,\boldsymbol{x}_{P-1}^T\right]^T$ an $MP\times 1$ vector with block entries of size $M\times 1$ each. Then, we obtain the following inequalities:

$$\|\boldsymbol{\Xi}\|_{b\infty} = \max_{\boldsymbol{x}\neq\boldsymbol{0}} \frac{\max\left(\sum_{p=0}^{P-1}\left\|\boldsymbol{\Xi}_p \boldsymbol{x}_p\right\|_2, \|\boldsymbol{x}_0\|_2, ..., \|\boldsymbol{x}_{P-2}\|_2\right)}{\|\boldsymbol{x}\|_{b\infty}} \le \max_{\boldsymbol{x}\neq\boldsymbol{0}} \frac{\max\left(\sum_{p=0}^{P-1}\left\|\boldsymbol{\Xi}_p \boldsymbol{x}_p\right\|_2, \|\boldsymbol{x}\|_{b\infty}\right)}{\|\boldsymbol{x}\|_{b\infty}}, \quad (33)$$

$$\begin{aligned}\sum_{p=0}^{P-1}\left\|\boldsymbol{\Xi}_p \boldsymbol{x}_p\right\|_2 &\le \left\|\boldsymbol{I}_M - \mu E\{\boldsymbol{A}(k)\}\right\|_2 \cdot \sum_{p=0}^{P-1}\left\|\beta_p \boldsymbol{x}_p\right\|_2 \\ &\le \left\|\boldsymbol{I}_M - \mu E\{\boldsymbol{A}(k)\}\right\|_2 \cdot \sum_{p=0}^{P-1}\beta_p \cdot \|\boldsymbol{x}\|_{b\infty} \\ &= \left\|\boldsymbol{I}_M - \mu E\{\boldsymbol{A}(k)\}\right\|_2 \cdot \|\boldsymbol{x}\|_{b\infty}.\end{aligned} \quad (34)$$

Inserting (34) into (33) yields

$$\|\boldsymbol{\Xi}\|_{b\infty} \le \max\left(\left\|\boldsymbol{I}_M - \mu E\{\boldsymbol{A}(k)\}\right\|_2, 1\right). \quad (35)$$

Since the spectral radius of a matrix is upper bounded by its any norm [30], it can be established that

$$\rho(\boldsymbol{\Xi}) \le \|\boldsymbol{\Xi}\|_{b\infty} \le \max\left(\left\|\boldsymbol{I}_M - \mu E\{\boldsymbol{A}(k)\}\right\|_2, 1\right) \le 1, \quad (36)$$

which leads further to

$$\rho\left(\boldsymbol{I}_M - \mu E\{\boldsymbol{A}(k)\}\right) < 1. \quad (37)$$

Equation (36) means that any eigenvalue $\lambda_j$ of $\boldsymbol{\Xi}$ satisfies $|\lambda_j| \le 1$ for $j = 1, ..., MP$. It is stressed that $\boldsymbol{\Xi}$ could have an eigenvalue $\lambda$ with $|\lambda| = 1$. However, (37) can remove this possibility. To prove it, we assume that such an eigenvalue exists, with an $MP\times 1$ eigenvalue vector $\boldsymbol{x}$ consisting of $\boldsymbol{x} = \left[\boldsymbol{x}_0^T, ..., \boldsymbol{x}_{P-1}^T\right]^T$. Also, again using (31), the following relation

$$\boldsymbol{\Xi}\boldsymbol{x} = e^{j\theta}\boldsymbol{x} \quad (38)$$

with respect to the angle $\theta$ can be expanded as

$$\left[\sum_{p=0}^{P-1}(\boldsymbol{\Xi}_p \boldsymbol{x}_p)^T, \boldsymbol{x}_0^T, ..., \boldsymbol{x}_{P-2}^T\right]^T = e^{j\theta}\left[\boldsymbol{x}_0^T, ..., \boldsymbol{x}_{P-1}^T\right]^T, \quad (39)$$

which further reduces to:

$$\left[\sum_{p=0}^{P-1}\boldsymbol{\Xi}_p e^{-j(p+1)\theta}\right]\boldsymbol{x}_{P-1} = \boldsymbol{x}_{P-1}. \quad (40)$$

By the triangular inequality of norms, we obtain

$$\begin{aligned}\left\|\sum_{p=0}^{P-1}\boldsymbol{\Xi}_p e^{-j(p+1)\theta}\right\|_2 &\le \sum_{p=0}^{P-1}\left\|\boldsymbol{\Xi}_p e^{-j(p+1)\theta}\right\|_2 \\ &\le \sum_{p=0}^{P-1}\left\|\boldsymbol{\Xi}_p\right\|_2 \\ &= \left(\sum_{p=0}^{P-1}\beta_p\right)\left\|\boldsymbol{I}_M - \mu E\{\boldsymbol{A}(k)\}\right\|_2 \\ &= \left\|\boldsymbol{I}_M - \mu E\{\boldsymbol{A}(k)\}\right\|_2 < 1.\end{aligned} \quad (41)$$

Since (42) makes the contradiction of the assumption $|\lambda| = 1$, we have $|\lambda_j| < 1$ for any eigenvalue $\lambda_j$ of $\boldsymbol{\Xi}$ [17]. Subsequently, by means of the eigenvalue decomposition of $E\{\boldsymbol{A}(k)\}$, then the range of the step size that guarantees the mean stability of the algorithm is obtained from (37):

$$0 < \mu < \frac{2}{\lambda_{\max}\left(E\{\boldsymbol{A}(k)\}\right)}. \quad (42)$$